\documentstyle[11pt,newpasp,twoside,epsf]{article}
\def\edcomment#1{\iffalse\marginpar{\raggedright\sl#1\/}\else\relax\fi}
\reversemarginpar

\begin{document}
\title{Recent Results from the Wide Angle Search for Planets (WASP)
Prototype}
\author{Stephen R. Kane$^1$, Keith Horne$^1$, Rachel A. Street$^2$, Don
L. Pollacco$^2$, David James$^3$, Yiannis Tsapras$^1$, Andrew Collier
Cameron$^1$}
\affil{$^1$School of Physics \& Astronomy, University of St Andrews, North
Haugh, St Andrews, Fife KY16 9SS, Scotland\\
$^2$APS Division, School of Physics, Queen's University of Belfast,
University Road, Belfast, BT7 1NN, Northern Ireland\\
$^3$Laboratoire d'Astrophysique, Observatoire de Grenoble, BP 53, F-38041,
Grenoble, Cedex 9, France}

\begin{abstract}
WASP0 is a prototype for what is intended to become a collection of WASPs
whose primary aim is to detect transiting extra-solar planets across the
face of their parent star. The WASP0 instrument is a wide-field (9-degree)
6.3cm aperture F/2.8 Apogee 10 CCD camera (2Kx2K chip, 16-arcsec pixels).
The camera is mounted piggy-back on a commercial 10-inch Meade telescope.
We present some recent results from the WASP camera, including observations
from La Palma of the known transiting planet around HD 209458 and
preliminary analysis of other stars located in the same field. We also
outline further problems which restrict the ability to achieve photon
limited precision with a wide-field commercial CCD.
\end{abstract}

\section{Introduction}

Of the indirect methods, the use of transits is rapidly developing into a
strong and viable means to detect extra-solar planets. A transit occurs
when the apparent brightness of a star decreases temporarily due to an
orbiting planet passing between the observer and the stellar disk. Since
this can only be observed when the orbital plane is approximately aligned
with the line of sight, this transit method clearly favours large planets
orbiting their parent stars at small orbital radii (``hot Jupiters'').
Hence, a large sample of stars must be monitored in order to detect
transiting planets.

In this report, we briefly describe the data reduction methods and initial
results from WASP0, a prototype wide-angle CCD camera currently being used
to search for transiting extra-solar planet signatures.

\section{Overview of WASP}

WASP0 is a prototype for what is intended to become a ``swarm'' of WASPs
whose primary aim is to detect transiting extra-solar planets. The WASP0
instrument is a wide-field (9-degree) 6.3cm aperture F/2.8 Nikon camera
lens, Apogee 10 CCD detector (2Kx2K chip, 16-arcsec pixels) which was
built by Don Pollaco at Queen's University, Belfast. The camera was
mounted piggy-back on a commercial 8-inch Celestron telescope during its
observing run at La Palma, Canary Islands from 20th June until 20th
August, 2000. It is currently mounted on a 10-inch Meade at Kryoneri,
Greece.

\section{Observations and Data Reduction}

In order to monitor a sufficient number of stars for a successful
planetary transit detection, a wide field needs to be combined with
reasonably crowded star fields. Field stars towards Draco were regularly
monitored for two months during the observations from La Palma in 2000.
More recent observations from Greece have targeted the Hyades open
cluster. During the first two months of WASP0 observations, nearly 150
Gbytes of data were obtained.

The data reduction pipeline currently being developed to reduce this
dataset uses a modified version of the photometry program DoPHOT
(Schechter, Mateo, \& Saha 1993). The positions of the stars on each frame
are ``fixed'' using a template image which reduces the number of PSF fit
parameters from 7 to 5, thus decreasing processing time and increasing the
quality of the photometry.

\begin{figure}
\plotfiddle{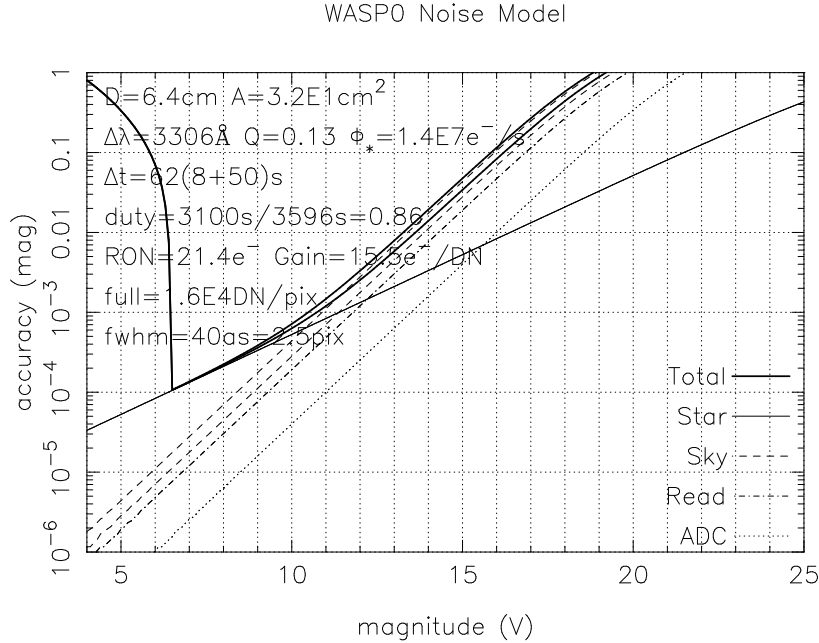}{8cm}{-90}{46}{46}{-170}{260}
\caption{Noise model for the WASP0 camera.}
\end{figure}

A transiting planet with $r \sim r_J$ and $M_\star<M_\odot$ may cause a
1-10\% decrease in brightness of the star. WASP0 aims to achieve milli-mag
photometry to unambiguously detect such effects. Signal-to-noise
calculations (see Figure 1) suggest that WASP0 will achieve 1\% rms
accuracy in 60 minutes for 13th-14th magnitude stars. This of course
depends upon the sky background, particularly associated with moonlight.

\section{Preliminary Results}

Figure 2 shows an RMS vs mag diagram for a single night of WASP0 data.
The solid line indicates the theoretical noise limit with the $\pm1\sigma$
range being shown by the dashed lines either side. The diagram contains
1500 stars from a $500\times500$ sub-frame of 320 images with exposure
times of 50 seconds. It can be seen that our present analysis of WASP0
images achieves the theoretical sky noise limited performance for
differential photometry, reaching 1\% rms accuracy at 11th magnitude.

\begin{figure}
\plotfiddle{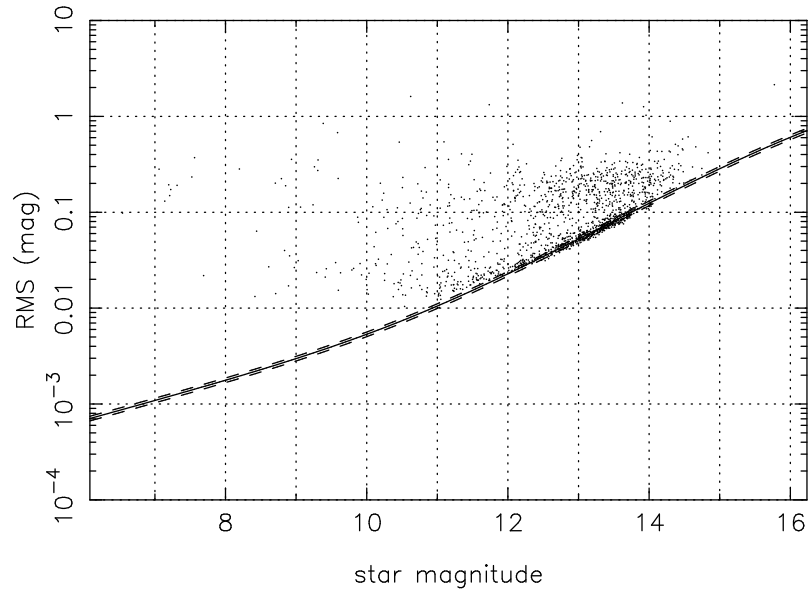}{7.2cm}{-90}{46}{46}{-170}{260}
\caption{Photometric accuracy vs star magnitude for 1500 stars in
comparison with theoretical accuracy predicted based on the CCD noise
model.}
\end{figure}

A known transiting extra-solar planet around HD 209458 (Charbonneau et
al. 2000; Henry et al. 2000) was observed using WASP0 at the predicted
transit time of 8th August, 2000. The top-left panel of Figure 3 clearly
shows the characteristic dip in the HD 209458 lightcurve due to the
planetary transit. The WASP0 data also contains a large collection of
variable star lightcurves. Shown in Figure 3 are three variables from the
same data.

\begin{figure}
\plotfiddle{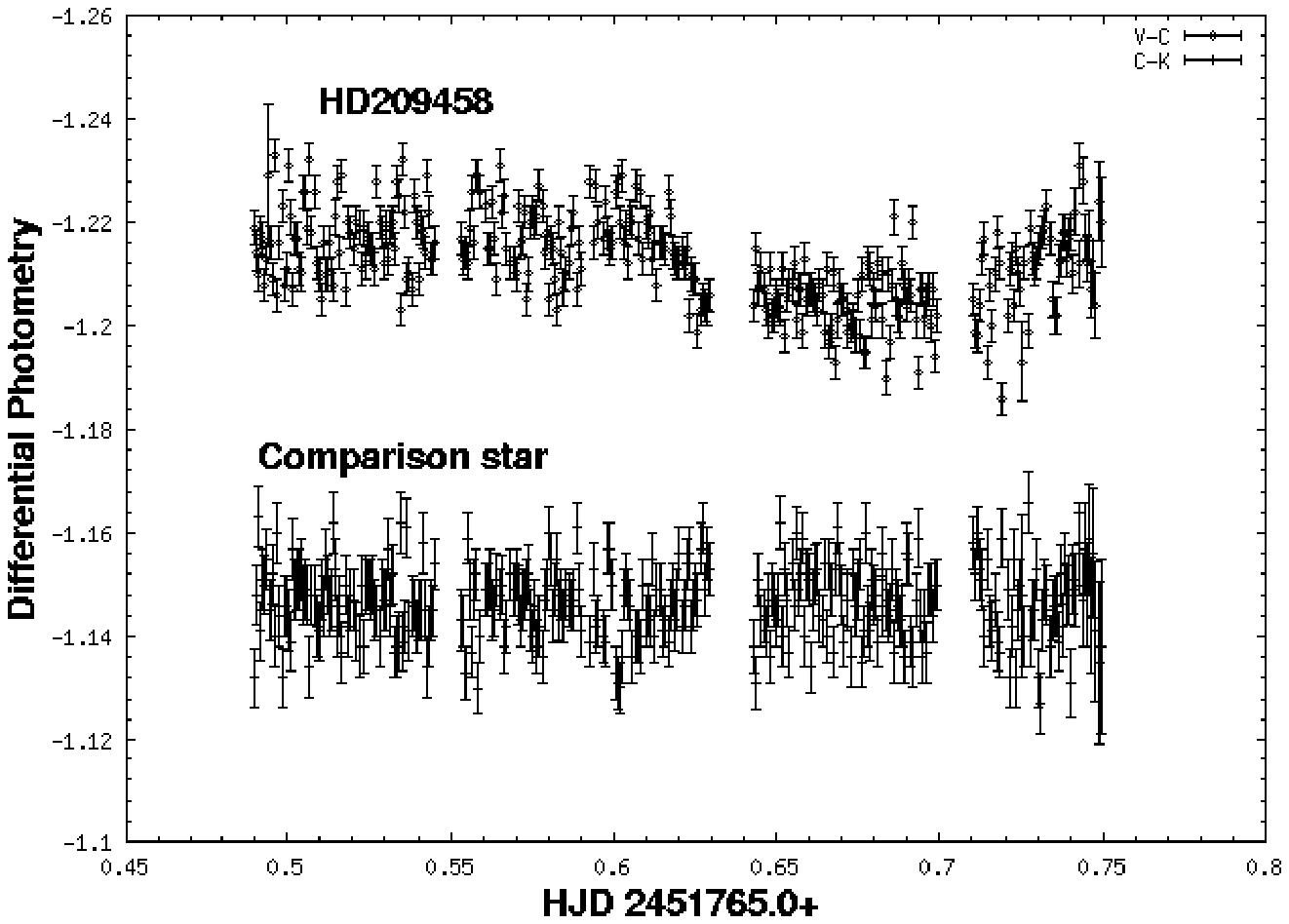}{6cm}{0}{41}{41}{-210}{-40}
\plotfiddle{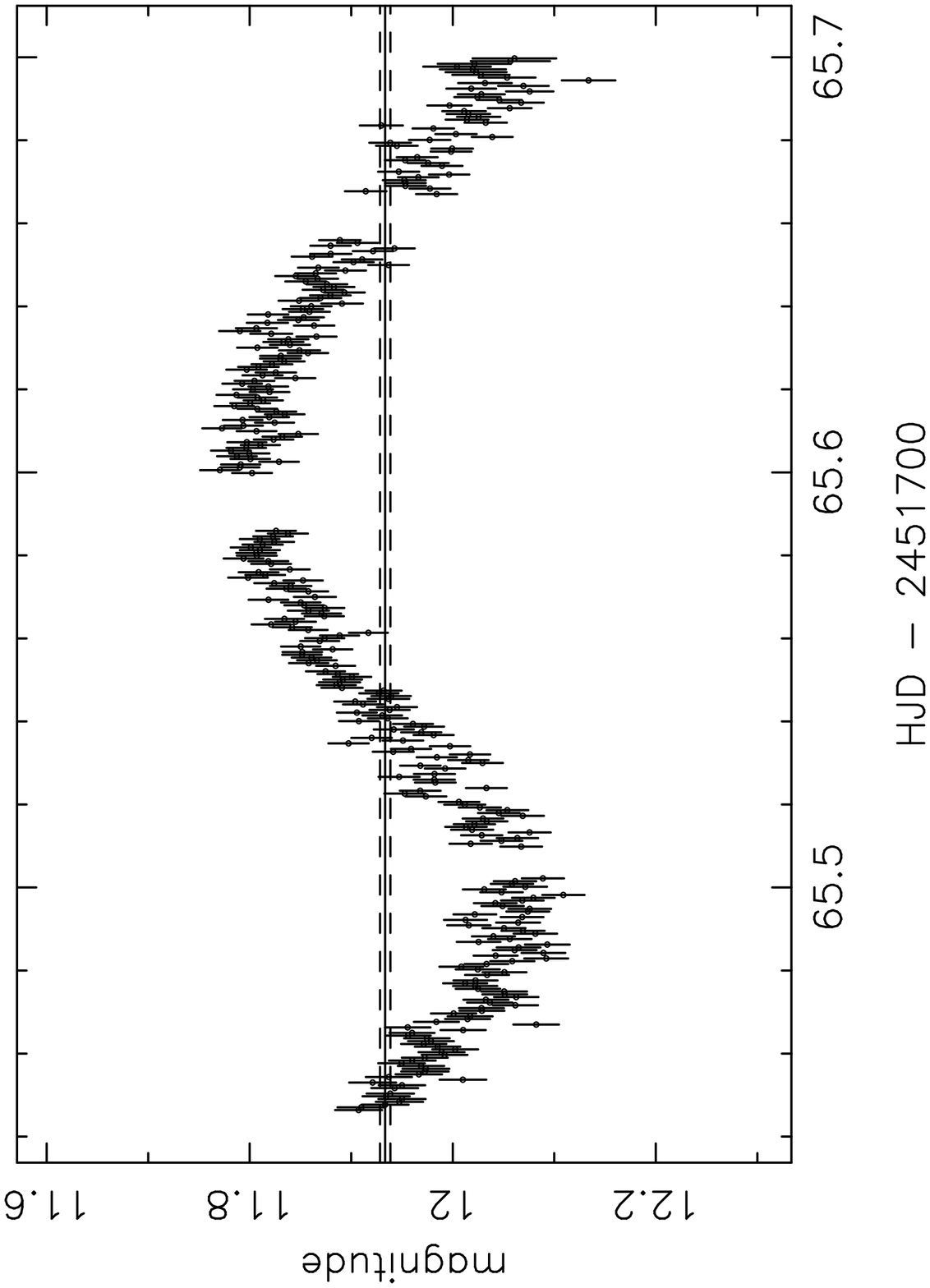}{0cm}{-90}{25}{25}{-10}{224}
\plotfiddle{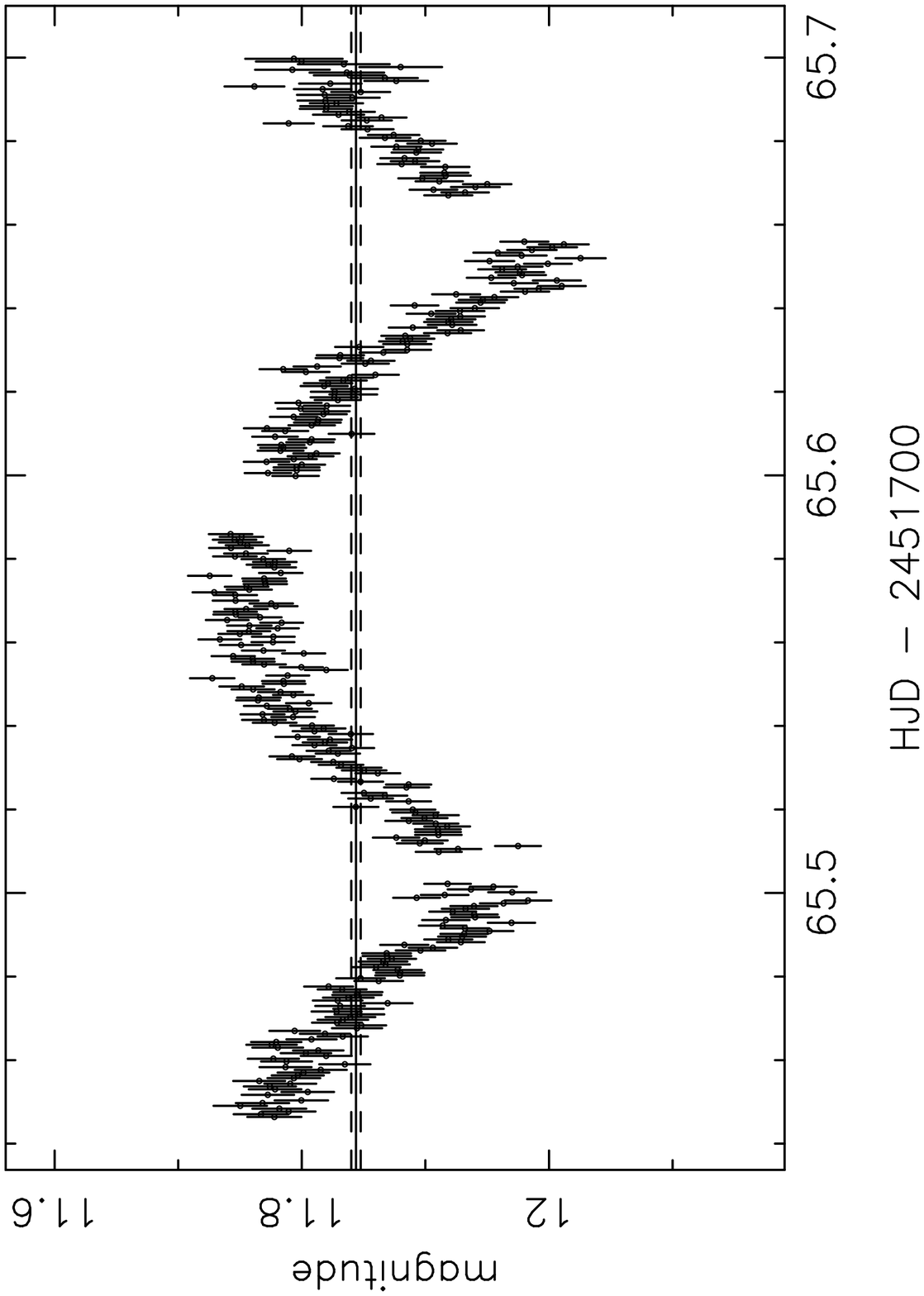}{0cm}{-90}{25}{25}{-180}{120}
\plotfiddle{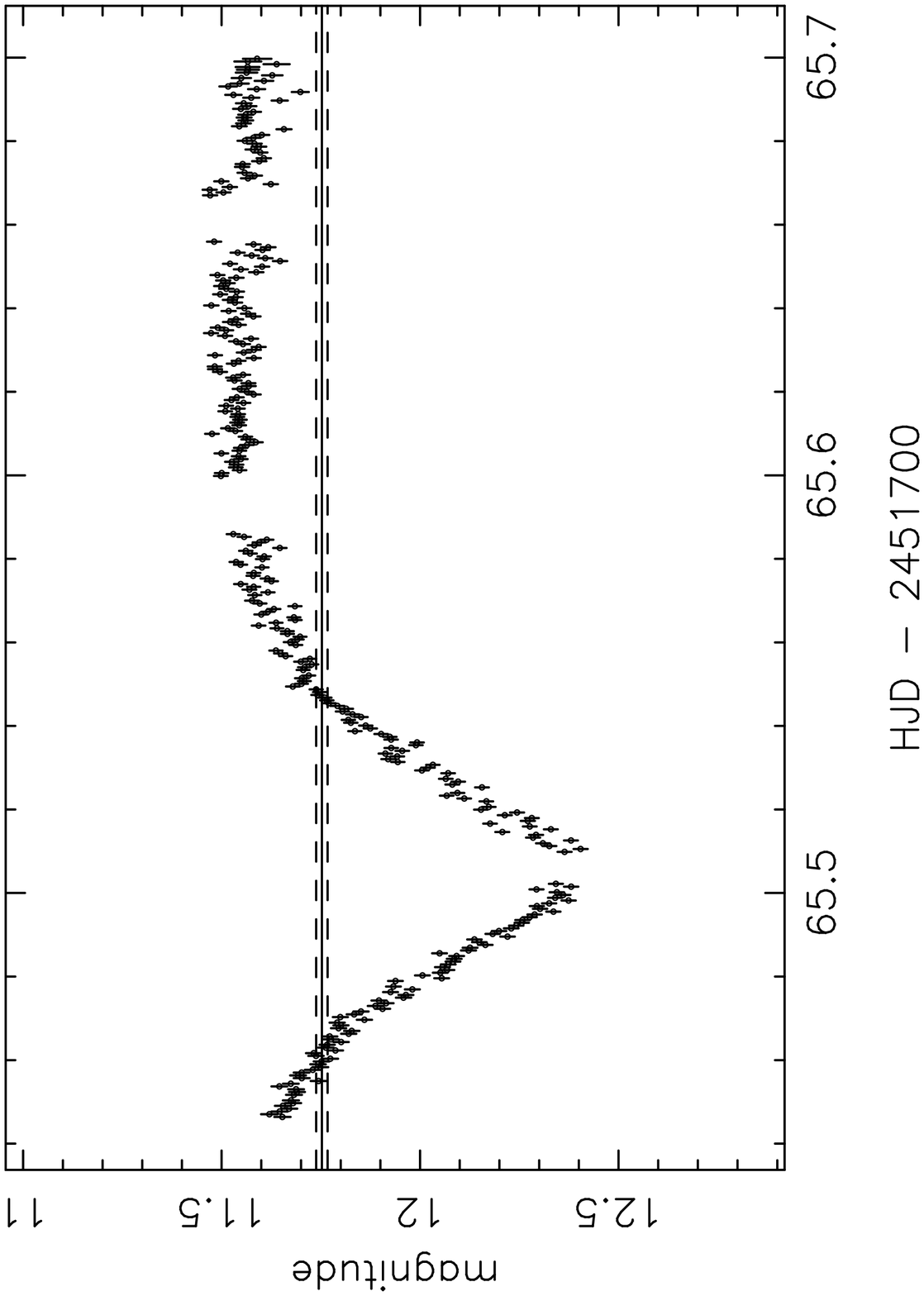}{0cm}{-90}{25}{25}{-10}{144}
\caption{Transit of HD 209458 and comparison star (top-left) along with
three variable stars from the the HD 209458 field}
\end{figure}

\section{Further Calibrations}

Cheap commercial CCD cameras may be used to detect extra-solar planetary
transits as long as systematic errors are treated with care. By using
thousands of lightcurves of constant stars available in the fields, the
systematic errors may be pushed below a milli-magnitude.

The adequate addressing of wide-field issues such as position dependent
airmass and vignetting is vital to achieve the required photometric
precision. Many of these factors are yet to be dealt with in this data,
including the construction of a PSF model which better matches the
undersampled stellar profiles. Star images with 1.7-pixel FWHM appear
4\% brighter when centred on a CCD pixel than when centred at the corner
between 4 pixels. The inclusion of an astrometric fit to fix star
positions and correct for field distortions is also being implemented.

\section{Summary and Future Work}

The preliminary analysis of WASP0 data presented here demonstrates that
this instrument is able to achieve the necessary precision required to
detect transit events due to extra-solar planets. Further calibrations
and refinement of the PSF model are needed to determine if WASP0 is able
to detect even smaller photometric deviations. This prototype has
successfully served as a proof-of-concept for SuperWASP. Further details
on the SuperWASP project are available in a separate paper by Street et
al. (2002) in this volume.

\end{document}